Title: Universalized Prisoner's Dilemma With Risk
Author: Paul Studtmann
Comments: 26 pages. Includes four pages of Mathematica worksheets that verify the mathematical claims made in the paper.
MSC-class: 91

Abstract: In this paper I present a mathematically novel approach to the Prisoner's Dilemma. I do so by first defining recursively a distinct action type, what I call 'universalizing', that I add to the original prisoner's dilemma. Such a modified version of the Prisoner's Dilemma provides a very good predictive model of the choices that would be made by agents in a prisoner's dilemma who trust each other. As I show, players playing a universalized prisoner's dilemma get as far out of the dilemma as is mathematically possible. I then add the concept of risk to the universalized version of prisoner's dilemma. Doing so provides a model that is sensitive to the trustworthiness of the agents in any prisoner's dilemma. As I show, with no risk, players manage to get out of the prisoner's dilemma; and with maximal risk, they succumb to it.

Consider two different scenarios both of which involve a situation that some philosophers and game theorists might want to model by way of the game known as the *Prisoner's Dilemma* (PD).  In scenario 1, Romeo and Juliet face a warden who gives them the following choices:  If each confesses, each will spend one minute in jail; if one confesses and the other doesn't, the one who confesses will spend fifty years in prison and the one who does not will go free; and if neither confesses, each will spend thirty years in prison.  In the second scenario, Valjean and Javert face a warden who gives them the same choices that he gave to Romeo and Juliet.

Although modeling both the above scenarios with PD may seem reasonable given the fact that both pairs of players face the same choices, it does not take too much reflection to see that the scenarios are significantly different and hence that appealing to PD as a model for both is problematic. To put the point bluntly: it is overwhelmingly likely that Romeo and Juliet, owing to their love for and hence trust of each other, will cooperate and overwhelmingly likely that Valjean and Javert, owing to their hatred and hence distrust of each other, will not.  As is well known, however, PD has only one Nash equilibrium: both parties confess.  Hence, although using PD to model Valjean and Javert may yield an intuitively correct result, using it to model Romeo and Juliet does not.

I will call any attempt to model situations like scenarios 1 and 2 with a game like PD *unresponsive.*  An unresponsive model is one that does not take into account the fact that the participants in a situation are responsive to the trustworthiness of the other people in the situation.  I take it as a datum that any unresponsive model seriously distorts our understanding of the dilemmas (and delights) that fill our lives.

In this paper, I propose a responsive model of situations like those typically modeled by PD. I do so in two stages. In the first stage, I propose another unresponsive model that lies at the furthest extreme from PD. I call such a model *Universalized Prisoner's Dilemma* (UPD). Two things will become clear about UPD. First, it is recursively defined in terms of PD. Second, UPD is a very good model of the scenario involving Romeo and Juliet but, not surprisingly given the fact that it is unresponsive, a very poor model of the scenario involving Valjean and Javert. In the second stage, I augment Universalized Prisoner's Dilemma by introducing risk. The result is what I call *Universalized Prisoner's Dilemma with Risk* (UPDR). Unlike PD, and unlike UPD, UPDR is a responsive game, one that can model both the scenario involving Romeo and Juliet and the scenario involving Valjean and Javert.

In section I of what follows, I provide informally a definition of the Universalizing operation. In section II, I apply the operation of Universalizing to the PD so as to yield UPD. I then go on to discuss the equilibrium solutions for UPD and show that UPD is indeed a very good model of Romeo and Juliet facing the choices described above. In section III, I attend to an objection that can be raised to introducing the universalizing operation as a way to model a prisoner's dilemma. In the course of my responding to the objection it shall become apparent that adding universalizing as an action type to symmetrical games has a number of theoretical virtues not the least of which is the fact that it provides a way to naturalize the concept of the strength of a moral obligation. Finally, in section IV, I add risk to UPD so as to yield *Universalized Prisoner's Dilemma with Risk* (UPDR).  I then go on to discuss the equilibrium solutions to UPDR so as to show that it adequately models both of the above scenarios.

Before proceeding, I should make one comment about the mathematics discussed in this paper.  As shall become clear, I shall present many mathematical claims without any proof.  The reason for the lack of proof is simple: most of the results obtained were obtained by computational software.  Hence, I could not provide the proofs even if I wanted to.  I have, however, included as an appendix Mathematica worksheets that verify all the mathematical claims being made.

*Section I – Universalizing*

To universalize a game requires adding to the initial action type in the game a second action type: universalizing. This introduces a somewhat unfortunate ambiguity with the term 'universalizing', though it is an ambiguity that should be easily resolved given any particular context. In the first sense, universalizing is an operation that can be performed on a game to yield another game. In the second sense, universalizing is a choice that can be made within a game that has been universalized in the first sense of that term.

When a player universalizes, he receives as a payoff what he would have received in the original game had everyone played the strategy he is playing in the universalized game. So if a player in UPD chooses to cooperate, if he also universalizes, he will receive the payoff he would have received in PD were the other player to cooperate as well. If, on the other hand, a player chooses to cooperate without universalizing, he will receive a payoff that depends on what the other player does: if the other player does not cooperate, the original player receives a payoff determined by PD under the assumption that he (the original player) cooperated and the other player did not; and if the other player does cooperate, then the original player receives a payoff determined by PD under the assumption that both players cooperated. Universalizing is thus a way to shield oneself from the effects of the other player's choice. A player who universalizes chooses, so to speak, to play PD among other like-minded players.

Before proceeding to an examination of the mathematical properties of UPD, it will be useful to discuss briefly the concept of universalizing that has just been introduced. One initial comment should be made: It is beyond doubt that the concept is mathematically coherent, since its definition is a completely standard recursive definition. The base case is given explicitly by PD. And the first and only time the operation is applied to PD, yields a universalized game. One could, were one so inclined, universalize again and again, ad infinitum. But for our purposes, one act of universalizing will do.

I emphasize the mathematical coherence of the universalizing relation to forestall an objection that might spring to mind. One might wonder just how players could in fact universalize in the real world. How does one get a payoff on the assumption that everyone plays the same action as you? After all, in many situations we face just the opposite problem: our interactions seem to have no effect on what others do. Such an objection, however, faces an immediate response based on the mathematical coherence of the universalizing operation. Given its coherence, the game it defines is coherent, and so one could rig up a mechanism to get players to play the game. For instance, one might incentivize a class of students with monetary rewards, give them the payoff schedule that is derived when PD is universalized, and watch them play the resulting game. No doubt, such students would eventually converge to the equilibrium strategies for UPD.

Of course, a universalized game would be more interesting were there to exist a natural mechanism that could cause (or incentivize if one wants to use that language) players to universalize. But of course there are many such mechanisms, for instance romantic love. That is why a universalized game is particularly appropriate for the study of a prisoner's dilemma involving Romeo and Juliet.

*Section II -- Universalized Prisoner's Dilemma*

When Z<Y<X<W, S stands for remain silent, and C stands for confess, the following is PD:

|   | S    | C    |
|---|------|------|
| S | X, X | Z, W |
| C | W, Z | Y, Y |

As is known, the only strategy in equilibrium in this game is the pure strategy of confession, C.

The following is the universalized version of PD ('U' stands for Universalize').

|      | SU   | CU   | S~U  | C~U  |
|------|------|------|------|------|
| SU   | X, X | X, Y | X, X | X, W |
| CU   | Y, X | Y, Y | Y, Z | Y, Y |
| S~U  | X, X | Z, Y | X, X | Z, W |
| C~U  | W, X | Y, Y | W, Z | Y, Y |

It may be of use to discuss some of the pairings to explain the values. Suppose both players play SU. Because they both universalize, they both receive as a payoff what they would have received in PD had both played S. Hence, they both receive X. Suppose player 1 plays CU and player 2 plays S~U. Because player 1 universalizes, he receives as a payoff what he would have received in PD had everyone played C. Hence, he receives Y. Player 2, on the other hand, does not universalize. Because he plays S, he receives what he would have received in PD had he played S and player 1 played what he is playing in the UPD, namely C. Hence player 2 receives Z. Suppose finally, that player 1 plays S~U and player 2 plays C~U. Because player 1 does not universalize, he receives what he would have received in PD had he played S and player 2 played what he is playing in UPD, namely C. Hence, player 1 receives Z. Because player 2 does not universalize, he receives what he would have received in PD had he played C and player 1 played what he is playing in UPD, namely S. Hence, player 2 receives W.

What, then, are the strategies in equilibrium in the Universalized game? There is only one. It is a mixed strategy equilibrium: SU/C~U. Hence, in the Universalized Game players will sometimes remain silent while universalizing and sometimes confess without universalizing. Let us allow q1 to represent the probability with which SU is played and q4 to represent the probability with which C~U is played. One can show that the following equations govern q1 and q4.

$$q1 = \frac{X - Y}{W - Y}$$

$$q4 = \frac{W - X}{W - Y}$$

To see what these equations mean for Romeo and Juliet, we can plug in the values from the scenario with which we began. In that scenario, X = 1 minute, W = 0 minutes, and Y = 30 years ≅ 11,500,000 minutes. Hence, q1 ≅ 11,499,999/12,500,000. Because q1 measures the probability that Romeo and Juliet will remain silent, we get the result that it is overwhelmingly likely that Romeo and Juliet will both remain silent.

In general, the equations governing q1 and q4 have the result that as the difference between X and Y increases so too does q1. The difference between X and

Y, however, is a measure of the seriousness of the dilemma facing the prisoners. Hence, just as one would expect between lovers, as the seriousness of the dilemma they face increases so too does the probability of cooperation. In fact, from the viewpoint of expected value, the rates of cooperation and non-cooperation are calibrated in a particularly interesting way. The expected value of playing the strategy SU/C~U is given by the following formula:

$$ev = q4W - q4^2W + X - q4X + q4^2Y$$

When this equation is solved simultaneously with the equation for q4, $ev$ reduces to the following:

$$ev = X$$

Expected value in UPD equals the expected value of PD were both players to remain silent. In other words, universalizing behavior gets players as far out of the prisoner's dilemma as is mathematically possible.

The behavior that emerges from UPD is thus much cheerier than the behavior that emerges from PD. Not only is a Pareto Optimal outcome achieved but cooperation increases as the cost of mutual defection increases. Nonetheless, the equilibrium solutions have a feature that keeps the behavior from being altogether cheery. The formulas for q1 and q4, which represent the probabilities with which players are silent or confess, contain only three variables: one for the outcome, X, that occurs under mutual silence in PD, one for the outcome, Y, that occurs under mutual confession in PD, and one for the reward, W, for confessing in PD when the other player remains silent. In a prisoner's dilemma, however, there is one other value at play: the punishment for remaining silent when the other person confesses.

Hence, the rate of confession is independent of this amount. So, the equilibrium solutions for UPD entail that some players (or all players some percent of the time) will engage in acts of non-cooperation in which they impose a potentially huge cost on a cooperating individual for a minimal gain. It does not take much of an imaginative stretch to see such actions as displays of *unbounded duplicity*. After all, some percentage of the population (or everyone some percentage of the time) will, if the opportunity arises, defect and thus condemn to death another person for even a two-minute reduction in a sentence.

UPD thus has a number of interesting results: players achieve the expected value of the strategy for mutual cooperation in PD, cooperate more as the cost of mutual defection increases, and engage in acts of unbounded duplicity some percentage of the time. Were one searching for a philosophical view that such a game nicely captures, there could be none more appropriate than Leibniz' view that this is the best of all possible worlds. Those playing UPD achieve an optimal outcome given the initial conditions and tend to cooperate more as the stakes of non-cooperation rise, despite the fact that UPD also entails acts that, were one to use a moral label, could accurately be described as extremely immoral. Such a world stands in stark contrast to the Hobbesian world that players of PD inhabit. Every player in PD inevitably and without fail defects and as a result finds himself in the worst possible outcome along with everyone else -- a shared hell of defectors.

*Section III – An Objection and a Reply*

The previous section shows that UPD accurately predicts the actions of lovers like Romeo and Juliet when they are faced with a jailer who gives them the choices that are typically modeled by way of PD.  Here, however, the claim that UPD models Romeo and Juliet under such a circumstance faces an objection. According to the objection, UPD fundamentally mischaracterizes the strategic situation in which Romeo and Juliet find themselves. The difficulty can be seen by noting that neither one can really choose to universalize.  Why?  Because universalizing entails that the player receives a payoff determined solely by his or her choice to remain silent or confess. So, suppose that Romeo remains silent and universalizes.  It is still possible for Juliet to confess. If so, then Romeo under the envisioned scenario would not get the payoff that would result from both players remaining silent.  Rather, he would get the harshest outcome while Juliet would go free.

In response to this objection, one must make a distinction between games that are representationally accurate and those that are predictively accurate. A representationally accurate game preserves exactly the choice structure and the values as they are presented in the scenario under question.  A predictively accurate model yields equilibria that are empirically adequate.  A representationally accurate model would require the game-theoretic model to look like PD.  The obvious difficulty with such a representationally accurate model is the fact that it gives wildly inaccurate predictions.  Hence, if game theoretic modeling of such scenarios is to be even remotely useful, something in the modeling apparatus must change. And there really are only three options.  First, one might change the values that are specified in the scenario.   So, for instance, one might argue that when the jailer presents Romeo and Juliet with their choices, they would both ascribe an extremely high disvalue to going free if the other remains incarcerated. One could then keep the existing choice structure and generate a plausible prediction about what they would choose. Second, one could preserve the choice structure and the values but alter in some way that mathematical apparatus used to generate the equilibria. Finally, one could retain the values and the Nash solution concept but alter the choice structure. Adding universalizing to PD is an instance of the latter approach. UPD is not meant to capture the choice structure of the situation that Romeo and Juliet face and so is not meant to be a representationally accurate model but is rather meant to be a predictively accurate model.

Given the interest of the prisoner's dilemma it is not surprising that each of these possible responses have been pursued, though only the latter two have been developed in a mathematically sophisticated way. Although this is not the place for a thorough discussion of the various alternatives, a brief discussion should make vivid by way of contrast the strategy employed in this paper. Ever since Axelrod's work on repeated games[1], philosophers, social scientists and mathematicians have found in repeated games a way to explain the emergence of cooperative behavior in an environment in which players are doomed to play the Prisoner's Dilemma. Although players who play the Prisoner's Dilemma once may fail to cooperate, players who play the Prisoner's Dilemma repeatedly may latch onto a more cooperative strategy like tit-for-tat. Unfortunately, the appeal to repeated games does not in any obvious way eliminate the Prisoner's Dilemma. Not only can non-cooperative strategies dominate, but Press and Dyson have shown such a problem can be quite acute in the face of what are called zero-determinant strategies.[2]

---

[1] Axelrod, Robert, *The Evolution of Cooperation*, New York: Basic Books, 1984.

In addition to repeated games, there are other attempts to get out of the prisoner's dilemma, which do not appeal to the repeated playing of games. Robert Aumann introduced the concept of a correlated game. A correlated equilibrium is a solution concept that is more general than Nash's equilibrium concept. In a correlated game, each payer chooses an action depending on his or her observation of a public signal.[3] Players playing a correlated game do get out of the prisoner's dilemma. It should be clear, however, that such a way of modeling the prisoner's dilemma changes the choice structure of the game, since players of a prisoner's dilemma do not get the advantage of a publicly observable signal. In this way, correlated games are like universalized games. Indeed, one might think of universalizing as a way of internalizing the public signal that correlated games rely on. However, while universalized games change the choice structure of the prisoner's dilemma, unlike correlated games they rely on the standard Nash equilibrium concept.

It is also possible to model the prisoner's dilemma with an asynchronous one-shot game. According to Skyrms, such an analysis of the prisoner's dilemma shows up in Hume.[4] In an asynchronous game, a player gets to move after observing what the other player has don. An asynchronous representation of the prisoner's dilemma, however, in addition to changing the choice structure of the situation does not successfully get players out of the dilemma.[5]

---

[2] Press, William and Dyson, Freeman. "Iterated Prisoner's Dilemma contains strategies that dominate any evolutionary opponent." *Proc Natl Acad Sci USA 109:10409–10413* (2012).
[3] Aumann, Robert (1974) "Subjectivity and correlation in randomized strategies." *Journal of Mathematical Economics* 1:67-96; (1987) "Correlated Equilibrium as an Expression of Bayesian Rationality. *Econometrica"* 55(1):1-18
[4] Skyrms, Brian (1998) "The Shadow of the Future," in Coleman and Morris (eds.), *Rational Commitment and Social Justice: Essays for Gregory Kavka*, New York, Cambridge University Press.

Perhaps the most interesting recent discovery concerning the prisoner's dilemma comes from physics. A number of recent papers have shown that game theory can be augmented so as to model the interactions of superposed and entangled particles.[6] Eisert, Wilkens, and Lewenstein have shown that although not all superposed particles get out of the prisoner's dilemma, entangled particles do. Universalizing turns out to be very much like entanglement. Entanglement, however, unlike universalizing presupposes the framework of superposition. Hence, physicists must change the understanding of a move in a game in order to model the concept of entanglement – whereas a move would typically be represented as a stochastic matrix, within quantum game theory a move is represented by a unitary matrix. Unlike entanglement universalizing can be represented by an ordinary stochastic matrix. Hence, although the concept of universalizing applies to entangled particles, ordinary biological, economic and moral agents can universalize. Moreover, like entangled particles, players who play a universalized game get out of the prisoner's dilemma.

Universalized games thus join a rich and varied approach to the prisoner's dilemma. Although one might thus think that universalized games are just one among many possible ways of looking at the prisoner's dilemma, there are a number of features of universalized games that make them stand out from the other approaches. The first thing to note about the operation of universalizing is that it is recursively defined. It hardly needs emphasizing that recursive definitions are a hugely powerful important and elegant type of definition. So moving to UPD is a move to a mathematical neighborhood that is populated by some of the most lavish mathematical mansions around. The same cannot be said for the other approaches to the prisoner's dilemma just discussed. The fact that universalizing is recursively defined opens up the possibility that other recursive definitions can be used so as to provide predictively accurate models for situations that have caused trouble for game theory.  If there were other such models, one could see adding universalizing not as an ad-hoc move but rather as part of a systematic approach to game theoretic contexts.

---

[5] Kuhn, Steven, "Prisoner's Dilemma", *The Stanford Encyclopedia of Philosophy* (Fall 2014 Edition), Edward N. Zalta (ed.), URL = <http://plato.stanford.edu/archives/fall2014/entries/prisoner-dilemma/>.

[6] David A. Meyer, Phys. Rev. Lett. 82 (1999) 1052–1055, *Quantum Strategies*. J. Eisert, M. Wilkens, M. Lewenstein, Phys. Rev. Lett. 83 (1999) 3077–3080, *Quantum Games and Quantum Strategies*, N. F. Johnson, *Playing a Quantum Game with a Corrupted Source*. L. Marinatto, T. Weber, Phys. Lett. A 272 (2000) 291-303, *A Quantum Approach To Static Games of Complete Information*. T. Cheon, I. Tsutsui, Phys. Lett. A 348 (2006) 147-152, *Classical and Quantum Contents of Solvable Game Theory on Hilbert Space*.

      A second virtue of adding universalizing to PD so as to yield a predictively accurate model stems from the fact that it is not just a recursive definition but plausibly the simplest recursively definable operation that yields a game in which the equilibrium strategy has an expected value that equals the expected value of full cooperation in PD. I say 'plausibly' because the operation of universalizing is really quite simple and intuitive.  Because the complexity of operations can be measured, and because there would only be a finite number of simpler operations, if universalizing is indeed the simplest such operation it would be provably the simplest.  Although philosophers may wrangle about the extent to which simplicity counts in favor of a theory, surely the fact, if it is indeed a fact, that universalizing is

the simplest recursive definition that gets players out of the negative equilibrium of PD is a particularly notable feature.

The previous considerations in favor of universalizing PD are strictly mathematical. There is, however, a very powerful philosophical consideration. Many philosophers have appealed to game theory as a way to understand or analyze various normative situations. Their appeals to game theory, however, have employed games that do not contain concepts that lie close to the heart of normativity. Instead, the games contain concepts like remaining silent or confessing, correlating behavior, and so on; and such games are then used to understand normative situations. Although such a project is not entirely without interest, a philosophically more satisfying approach to analyzing normativity via game theory is to incorporate concepts that are very central to normativity into games themselves. UPD does just that. Ever since Kant, the importance of the concept of universalizing has been recognized as central to normative concerns. Although there may be other ways to understand the concept of universalizing, adding it as a recursively defined action type to game theory, in addition to having the mathematical virtues just discussed, yields a powerful game theoretic model of morality.

UPD is a symmetrical universalized game. Such games provide a very fruitful way of analyzing the strength of moral obligations. The key to such an analysis lies in interpreting the probabilities that govern the mixed strategies as such a measure. Let us suppose that the morally correct course of action in a prisoner's dilemma is to remain silent. Such a supposition does not in itself say how strong the moral obligation is. But the probability that someone remains silent in UPD can tell one how strong the obligation is:

$$q1 = \frac{X - Y}{W - Y}$$

As a measure of the strength of a moral obligation, this formula seems exactly right. As the disvalue of confessing grows relative to the value of remaining silent, the moral obligation to remain silent increases. In a prisoner's dilemma, then, as the stakes grow, so too does the moral obligation to cooperate. Of course, as the stakes grow so too does the temptation of a non-moral person to confess. But as UPD shows, agents who are disposed to act morally to each other will not succumb to such a temptation. Rather, the probability of their cooperation increases in step with the moral obligation to cooperate. This is not to say that morally inclined agents always act morally. As already discussed, there will always be some non-zero probability in UPD that an agent confesses. Even lovers who are as passionately committed to each other as Romeo and Juliet will with some predictable probability engage in duplicitous behavior.[7]

We began with an attempt to find a predictively adequate model of the behavior of agents like Romeo and Juliet in situations that are typically modeled by PD. By universalizing PD, we arrived not just at a game that is predictively accurate but also at a way to incorporate a fundamental normative concept into game theory.

---

[7] This analysis of universalizing should go some way toward rebutting the criticisms that Kenneth Binmore makes of the concept. Binmore, Kenneth (1994). *Game Theory and the Social Contract*: Volume 1: *Playing Fair*. Cambridge: MIT Press, pp 300-304.

The result is a model that can serve two functions simultaneously: it not only predicts behavior but it also measures the strength of the moral obligation that occurs in the situation in which the behavior occurs. Such games can thus be considered a way to naturalize at least part of the moral domain. Symmetrical universalized games, in addition to occurring within a purely extensional mathematical framework, and in addition to being the simplest recursively definable games that gets players out of the prisoner's dilemma, not only model physical behavior but also contain the resources to define at least one fundamental normative concept, namely the strength of a moral obligation.

Despite discovering such a remarkably fecund theoretical apparatus, however, we have not yet found what we have set out to find. The goal of this paper was to find a responsive model of a prisoner's dilemma. What we now have are two unresponsive games: PD and UPD. One might of course take each separately and apply it to whichever situations one sees fit. But it would be much more satisfying were there a single overarching model that had PD and UPD as limiting cases joined by a dimension that has not yet been added to the games.

*Section IV – Universalized Prisoner's Dilemma With Risk*

The means for the unification of PD and UPD can be seen by considering what separates the two scenarios with which we began. The most obvious difference between the two situations is the trustworthiness of the other player: Valjean would never trust Javert to remain silent if and only if Valjean does; whereas Romeo would trust Juliet completely to remain silent if and only if he does. If I find my counterpart to be particularly untrustworthy, then playing a universalized strategy would be foolish. If the other person is not trustworthy, playing a universalized strategy will cause one to remain silent when in fact one should have confessed. Although universalizing may come with a reward, like all things in life, the possibility of reward inevitably comes with risk. Hence, a very natural way to define a model that combines PD and UPD is to allow players to universalize but to modify the effect of their doing so by the risk involved.

To introduce risk into the model, we can augment UPD with a variable r. Maximal risk occurs when universalizing is entirely ineffective. In such a case, the absolute value of r would equal the difference between the outcomes of mutual cooperation and mutual defection in PD, i.e. X-Y. In a prisoner's dilemma, there is no risk associated with universalizing and confessing: If the other player does not confess, then the original player is even better off than he would have been had the other person confessed. Universalizing when remaining silent, however, does come with significant risk. Intuitively, the amount of risk is inversely proportional to the degree of trustworthiness of one's partner. Hence, in order to augment UPD with risk, we must add –r to the row in which player 1 is silent and universalizes and the column in which player 2 is silent and universalizes. The following game, *Universalized Prisoner's Dilemma with Risk* (UPDR), is the result:

|  | S/U | C/U | S/~U | C/~U |
|--|-----|-----|------|------|
|  |     |     |      |      |

| S/U | X-r,X-r | X-r,Y | X-r,X | X-r,Z |
| C/U | Y,X-r | Y,Y | Y,W | Y,Y |
| S/~U | X,X-r | Z,Y | X,X | Z,W |
| C/~U | W,X-r | Y,Y | W,Z | Y,Y |

What, then, is the effect of adding risk to UPD? Well, it should be easy enough to see from this payoff schedule that when risk is minimal, i.e. r=0, UPDR reduces to UPD. Hence, Romeo and Juliet, when playing this game, can each correctly make the assessment that the other poses no risk and end up playing UPD.

And what about those cases when risk is between zero and the maximal level? As it turns out, when $0 \leq r < X - Y$, the only strategy in equilibrium is the same mixed strategy that is in equilibrium in UPD: SU/C~U. Moreover, the following formulas govern the rates of silence and confession respectively:

$$q1 = \frac{X - Y - r}{W - Y}$$

$$q4 = \frac{W - X + r}{W - Y}$$

One can see here explicitly what was easily seen in the above payoff schedule: under minimal risk, i.e. when r=0, the rates of silence and confession reduce to the rates of silence and confession for UPD. And what about when risk is maximal, i.e. when r=X-Y? Under such a condition, there are three strategies in equilibrium: the pure strategy of confess and don't universalize; the pure strategy of confess and universalize; and the mixed strategy of confess and universalize and confess and don't universalize. Hence, when risk is maximal, prisoner's always confess, which of course is the only strategy in equilibrium in PD.

The expected value for playing SU/C~U is given by the following formula:

$$ev = -r + q4r + q4W - q4^2W + X - q4X + q4^2Y$$

Solving this formula simultaneously with the formula for q4 above yields the following formula:

$$ev = X - r$$

Hence, as risk approaches the maximal level, i.e. as r approaches $X - Y$, the expected value of playing UPDR approaches Y, which is the expected value of mutual defection in PD. Under maximal risk, therefore, UPDR reduces both in strategy and in expected value to PD. And when risk is between zero and the maximal risk, the rates of confession and silence as well as the expected value of playing the various strategies fall between what would happen in UPD and what would happen in PD.

UPDR is thus a responsive model of a prisoner's dilemma. Moreover, it is one that, unlike either of its unresponsive counterparts, can provide an accurate, indeed truistic, assessment of our existential situation: The world we inhabit is somewhere between the best and worst of all possible worlds; and just how it goes for us in this world depends on the trustworthiness of those with whom we play the games that occupy our lives.

# Appendix

In the following Mathematica worksheet, I first compute the equilibrium strategy for UPD. As one can see, this involves computing the solutions to 15 sets of equations. I use the following notation – 'i' stands for interact; 'u' stands for universalize; 'n' stands for does not. I place the 'n' after the term that it negates. So, for instance, 'inun' means does not interact and does not universalize. There is only one set that has a solution and hence there is only one strategy in equilibrium. After solving the fifteen sets of equations, I then compute the expected value of playing the strategy in equilibrium.

```
ClearAll[su, cu, snu, cnu, q1, q2, q3, q4, ev]
```

In[80]:= 
```
su = q1 * X + q2 * X + q3 * X + q4 * X
cu = q1 * Y + q2 * Y + q3 * Y + q4 * Y
snu = q1 * X + q2 * Z + q3 * X + q4 * Z
cnu = q1 * W + q2 * Y + q3 * W + q4 * Y
```

Out[80]= q1 X + q2 X + q3 X + q4 X

Out[81]= q1 Y + q2 Y + q3 Y + q4 Y

Out[82]= q1 X + q3 X + q2 Z + q4 Z

Out[83]= q1 W + q3 W + q2 Y + q4 Y

In[84]:= 
```
Reduce[{su == cu, su ≥ snu, su ≥ cnu, cu ≥ snu, cu ≥ snu,
    q1 + q2 == 1, q1 > 0, q2 > 0, q4 == 0, q3 == 0, Z < Y, Y < X, X < W}]
```

Out[84]= False

In[85]:= 
```
Reduce[{su == snu, su ≥ cu, su ≥ cnu, snu ≥ cu, snu ≥ cnu,
    q1 + q3 == 1, q1 > 0, q3 > 0, q2 == 0, q4 == 0, Z < Y, Y < X, X < W}]
```

Out[85]= False

In[86]:= 
```
Reduce[{su == cnu, su ≥ cu, su ≥ snu, cnu ≥ cu, cnu ≥ snu,
    q1 + q4 == 1, q1 > 0, q4 > 0, q2 == 0, q3 == 0, Z < Y, Y < X, X < W}]
```

Out[86]= Z ∈ Reals && Y > Z && W > Y && 0 < q4 < 1 && X == W − q4 W + q4 Y && q3 == 0 && q2 == 0 && q1 == $\frac{q4\, X - q4\, Y}{W - X}$

In[87]:= 
```
Reduce[{cu == snu, cu ≥ su, cu ≥ cnu, snu ≥ su, snu ≥ cnu,
    q2 + q3 == 1, q1 == 0, q3 > 0, q2 > 0, q4 == 0, Z < Y, Y < X, X < W}]
```

Out[87]= False

In[88]:= 
```
Reduce[{cu == cnu, cu ≥ su, cu ≥ snu, cnu ≥ su, cnu ≥ snu,
    q2 + q4 == 1, q1 == 0, q2 > 0, q4 > 0, q3 == 0, Z < Y, Y < X, X < W}]
```

Out[88]= False

In[89]:= 
```
Reduce[{snu == cnu, snu ≥ su, snu ≥ cu, cnu ≥ su, cnu ≥ cu,
    q3 + q4 == 1, q1 == 0, q2 == 0, q3 > 0, q4 > 0, Z < Y, Y < X, X < W}]
```

Out[89]= False

In[90]:= 
```
Reduce[{su == cu, su == snu, cu == snu, su ≥ cnu, cu ≥ cnu,
    snu ≥ cnu, q1 + q2 + q3 == 1, q4 == 0, q1 > 0, q2 > 0, q3 > 0, Z < Y, Y < X, X < W}]
```

Out[90]= False

In[91]:= 
```
Reduce[{su == cu, su == cnu, cu == cnu, su ≥ snu, cu ≥ snu,
    cnu ≥ snu, q1 + q2 + q4 == 1, q3 == 0, q1 > 0, q2 > 0, q4 > 0, Z < Y, Y < X, X < W}]
```

Out[91]= False

```mathematica
In[92]:= Reduce[{su == snu, su == cnu, snu == cnu, su ≥ cu, snu ≥ cu,
    cnu ≥ cu, q1 + q3 + q4 == 1, q2 == 0, q1 > 0, q3 > 0, q4 > 0, Z < Y, Y < X, X < W}]

Out[92]= False

In[93]:= Reduce[{cu == snu, cu == cnu, snu == cnu, cu ≥ su, snu ≥ su,
    cnu ≥ su, q2 + q3 + q4 == 1, q1 == 0, q2 > 0, q3 > 0, q4 > 0, Z < Y, Y < X, X < W}]

Out[93]= False

In[94]:= Reduce[{cu == snu, cu == cnu, snu == cnu, cu == su, snu == su, cnu == su,
    q1 + q2 + q3 + q4 == 1, q1 > 0, q2 > 0, q3 > 0, q4 > 0, Z < Y, Y < X, X < W}]

Out[94]= False

In[95]:= Reduce[{su ≥ cu, su ≥ snu, su ≥ cnu, q1 == 1, q2 == 0, q3 == 0, q4 == 0, Z < Y, Y < X, X < W}]

Out[95]= False

In[96]:= Reduce[{cu ≥ su, cu ≥ snu, cu ≥ cnu, q2 == 1, q1 == 0, q3 == 0, q4 == 0, Z < Y, Y < X, X < W}]

Out[96]= False

In[97]:= Reduce[{snu ≥ su, snu ≥ cu, snu ≥ cnu, q3 == 1, q2 == 0, q3 == 0, q4 == 0, Z < Y, Y < X, X < W}]

Out[97]= False

In[98]:= Reduce[{cnu ≥ su, cnu ≥ cu, cnu ≥ snu, q4 == 1, q2 == 0, q3 == 0, q1 == 0, Z < Y, Y < X, X < W}]

Out[98]= False

In[99]:= Reduce[{ev == (q1 * su + q4 * cnu), q2 == 0, q3 == 0, q1 + q4 == 1}, ev]

Out[99]= q3 == 0 && q2 == 0 && q1 == 1 - q4 && ev == q4 W - q4^2 W + X - q4 X + q4^2 Y

In[100]:= Reduce[{ev == q4 W - q4^2 W + X - q4 X + q4^2 Y, q1 == (q4 X - q4 Y)/(W - X), q1 + q4 == 1}]

Out[100]= W - Y ≠ 0 && q4 == (W - X)/(W - Y) && q1 == 1 - q4 && ev == X && W - X ≠ 0
```

In this Mathematica worksheet, I first compute the equilibrium strategies for UPDR. As one can see, there are five sets that have solutions. However, one of those sets, i.e. the one in which q1=1, requires there to be negative risk and so is not of interest. Three of those sets have solutions that require risk to be greater than or equal to X. Hence, they are limiting cases of risk, i.e. when risk is maximal. After computing the solution sets, I compute the expected value of the one strategy that is in equilibrium when risk is between 0 and X.

```
ClearAll[su, cu, snu, cnu, q1, q2, q3, q4, ev]
```
In[125]:=
```
su = q1 * (X - r) + q2 * (X - r) + q3 * (X - r) + q4 * (X - r)
cu = q1 * Y + q2 * Y + q3 * Y + q4 * Y
snu = q1 * X + q2 * Z + q3 * X + q4 * Z
cnu = q1 * W + q2 * Y + q3 * W + q4 * Y
```

Out[125]= q1 (-r + X) + q2 (-r + X) + q3 (-r + X) + q4 (-r + X)

Out[126]= q1 Y + q2 Y + q3 Y + q4 Y

Out[127]= q1 X + q3 X + q2 Z + q4 Z

Out[128]= q1 W + q3 W + q2 Y + q4 Y

In[129]:=
```
Reduce[{su == cu, su ≥ snu, su ≥ cnu, cu ≥ snu, cu ≥ snu,
   q1 + q2 == 1, q1 > 0, q2 > 0, q4 == 0, q3 == 0, Z < Y, Y < X, X < W}]
```
Out[129]= False

In[130]:=
```
Reduce[{su == snu, su ≥ cu, su ≥ cnu, snu ≥ cu, snu ≥ cnu,
   q1 + q3 == 1, q1 > 0, q3 > 0, q2 == 0, q4 == 0, Z < Y, Y < X, X < W}]
```
Out[130]= False

In[131]:=
```
Reduce[{su == cnu, su ≥ cu, su ≥ snu, cnu ≥ cu, cnu ≥ snu,
   q1 + q4 == 1, q1 > 0, q4 > 0, q2 == 0, q3 == 0, Z < Y, Y < X, X < W}]
```
Out[131]= Y ∈ Reals && X > Y && W > X && Z < Y && 0 < q4 < 1 &&
  r == -W + q4 W + X - q4 Y && q3 == 0 && q2 == 0 && q1 == $\frac{-q4\,r + q4\,X - q4\,Y}{r + W - X}$

In[132]:=
```
Reduce[{cu == snu, cu ≥ su, cu ≥ cnu, snu ≥ su, snu ≥ cnu,
   q2 + q3 == 1, q1 == 0, q3 > 0, q2 > 0, q4 == 0, Z < Y, Y < X, X < W}]
```
Out[132]= False

In[133]:=
```
Reduce[{cu == cnu, cu ≥ su, cu ≥ snu, cnu ≥ su, cnu ≥ snu,
   q2 + q4 == 1, q1 == 0, q2 > 0, q4 > 0, q3 == 0, Z < Y, Y < X, X < W}]
```
Out[133]= Y ∈ Reals && X > Y && W > X && r ≥ X - Y && 0 < q4 < 1 && Z < Y && q3 == 0 && q1 == 0 && q2 == 1 - q4

In[134]:=
```
Reduce[{snu == cnu, snu ≥ su, snu ≥ cu, cnu ≥ su, cnu ≥ cu,
   q3 + q4 == 1, q1 == 0, q2 == 0, q3 > 0, q4 > 0, Z < Y, Y < X, X < W}]
```
Out[134]= False

In[135]:=
```
Reduce[{su == cu, su == snu, cu == snu, su ≥ cnu, cu ≥ cnu,
   snu ≥ cnu, q1 + q2 + q3 == 1, q4 == 0, q1 > 0, q2 > 0, q3 > 0, Z < Y, Y < X, X < W}]
```
Out[135]= False

In[136]:=
```
Reduce[{su == cu, su == cnu, cu == cnu, su ≥ snu, cu ≥ snu,
   cnu ≥ snu, q1 + q2 + q4 == 1, q3 == 0, q1 > 0, q2 > 0, q4 > 0, Z < Y, Y < X, X < W}]
```
Out[136]= False

In[137]:= **Reduce[{su == snu, su == cnu, snu == cnu, su ≥ cu, snu ≥ cu,
  cnu ≥ cu, q1 + q3 + q4 == 1, q2 == 0, q1 > 0, q3 > 0, q4 > 0, Z < Y, Y < X, X < W}]**

Out[137]= False

In[138]:= **Reduce[{cu == snu, cu == cnu, snu == cnu, cu ≥ su, snu ≥ su,
  cnu ≥ su, q2 + q3 + q4 == 1, q1 == 0, q2 > 0, q3 > 0, q4 > 0, Z < Y, Y < X, X < W}]**

Out[138]= False

In[139]:= **Reduce[{cu == snu, cu == cnu, snu == cnu, cu == su, snu == su, cnu == su,
  q1 + q2 + q3 + q4 == 1, q1 > 0, q2 > 0, q3 > 0, q4 > 0, Z < Y, Y < X, X < W}]**

Out[139]= False

In[140]:= **Reduce[{su ≥ cu, su ≥ snu, su ≥ cnu, q1 == 1, q2 == 0, q3 == 0, q4 == 0, Z < Y, Y < X, X < W}]**

Out[140]= X ∈ Reals && W > X && r ≤ -W + X && Y < X && Z < Y && q4 == 0 && q3 == 0 && q2 == 0 && q1 == 1

In[141]:= **Reduce[{cu ≥ su, cu ≥ snu, cu ≥ cnu, q2 == 1, q1 == 0, q3 == 0, q4 == 0, Z < Y, Y < X, X < W}]**

Out[141]= Z ∈ Reals && Y > Z && X > Y && W > X && r ≥ X - Y && q4 == 0 && q3 == 0 && q2 == 1 && q1 == 0

In[142]:= **Reduce[{snu ≥ su, snu ≥ cu, snu ≥ cnu, q3 == 1, q2 == 0, q3 == 0, q4 == 0, Z < Y, Y < X, X < W}]**

Out[142]= False

In[143]:= **Reduce[{cnu ≥ su, cnu ≥ cu, cnu ≥ snu, q4 == 1, q2 == 0, q3 == 0, q1 == 0, Z < Y, Y < X, X < W}]**

Out[143]= Z ∈ Reals && Y > Z && X > Y && W > X && r ≥ X - Y && q4 == 1 && q3 == 0 && q2 == 0 && q1 == 0

In[144]:= **Reduce[{ev == (q1 * su + q4 * cnu), q2 == 0, q3 == 0, q1 + q4 == 1}, ev]**

Out[144]= q3 == 0 && q2 == 0 && q1 == 1 - q4 && ev == -r + q4 r + q4 W - q4$^2$ W + X - q4 X + q4$^2$ Y

In[145]:= **Reduce[{ev == -r + q4 r + q4 W - q4$^2$ W + X - q4 X + q4$^2$ Y, q1 == $\frac{-q4\,r + q4\,X - q4\,Y}{r + W - X}$, q1 + q4 == 1}]**

Out[145]= W - Y ≠ 0 && q4 == $\frac{r + W - X}{W - Y}$ && q1 == 1 - q4 && ev == -r + X && r + W - X ≠ 0